%% file: Letter_PRLv3.tex
\documentclass[prl,twocolumn,superscriptaddress,nofootinbib,aps,10pt,preprintnumbers,
longbibliography]{revtex4-1}

\usepackage[colorlinks=true,linkcolor=black,citecolor=blue,urlcolor=blue, pdfborder={0 0 0}]{hyperref}
 \usepackage{amssymb,amsmath,amsthm,cancel,graphicx}
\usepackage{mathtools}
\usepackage[utf8]{inputenc}
\usepackage{color}
\usepackage[table,xcdraw,dvipsnames]{xcolor}
\usepackage{multirow}
\usepackage[normalem]{ulem}    

\newcommand{\Sec}[1]{ \smallskip \noindent {\sl \bfseries #1}}

\definecolor{ultramarine}{RGB}{0,32,96}

\def\Tm{T^m}
\def\mFb{{\mathcal F}_b}
\def\mP{m_{\rm P}}

\def\mR{\mathcal{R}}

\def\zlc{z_{\rm LC}}
\def\tlc{t_{\rm LC}}
\def\taulc{\tau_{\scriptscriptstyle\rm LC}}
\def\Tlc{T_{\rm LC}}

\def\zde{z_{\rm DE}}
\def\Tde{T_{\rm DE}}
\def\tde{t_{\rm DE}}
\def\Rde{R_{\rm DE}}

\def\eq#1{{Eq.~(\ref{#1})}}
\def\eqs#1#2{{Eqs.~(\ref{#1})-(\ref{#2})}}

\newcommand{\gsim}{\gtrsim}

\def\beq{\begin{equation}}  
\def\eeq{\end{equation}}
\def\beqa{\begin{eqnarray}}  
\def\eeqa{\end{eqnarray}}
\def\({\left(}
\def\){\right)}
\def\[{\left[}
\def\]{\right]}

\def\varv{v}

\def\vev#1{\left\langle #1\right\rangle}

\def\tr{\mbox{Tr}\,}

\def\diag{\mbox{diag}\,}

\def\sq{\nu}

\renewcommand{\bar}{\overline}

\newcommand{\mM}{\mathcal{M}}

\begin{document}

\title{
Accelerated cosmic expansion, mass creation, and the 
QCD axion}

\newcommand{\LNF}{\small \it INFN, Laboratori Nazionali di Frascati, C.P.~13, 100044 Frascati, Italy}
\newcommand{\NICPB}{\small \it Laboratory of High Energy and Computational Physics, NICPB,  R\"avala 10, 10143 Tallinn, Estonia }

\newcommand{\SISSA}{\small \it SISSA International School for Advanced Studies, Via Bonomea 265, 34136, Trieste, Italy}
\newcommand{\INFN}{\small \it INFN,  Sezione di Trieste, Via Bonomea 265, 34136, Trieste, Italy}

 \author{Kristjan M\"u\"ursepp}
\email{kristjan.muursepp@kbfi.ee}
\affiliation{\NICPB}
\affiliation{\LNF}

 \author{Enrico Nardi}
\email{enrico.nardi@lnf.infn.it}
\affiliation{\NICPB}
\affiliation{\LNF}

\author{Clemente Smarra}
\email{csmarra@sissa.it\\ }
\affiliation{\SISSA}\affiliation{\INFN}

\begin{abstract}
We propose a mechanism in which the current acceleration of cosmic 
expansion is driven by continuous creation of energy density $\rho_b$ for a certain 
 field $\varphi_b$.  We   accordingly modify Einstein’s equation, 
derive modified Friedmann equations and analyze the regimes in which 
cosmic acceleration occurs. 
The creation process requires $\rho_b\neq 0$
as initial condition, which we enforce by 
identifying $\varphi_b$ with the axion of a hidden gauge group that 
confined in recent cosmological times, leading to a level crossing between $\varphi_b$ and the QCD axion, which is   
 assumed to comprise dark matter.  The 
 conversion  of a small fraction 
of QCD axions into $\varphi_b$  shortly before matter-dark energy equality  generates the  
 initial $\rho_b$ needed to trigger the creation process and offers a solution to the coincidence puzzle. 
\end{abstract}

\maketitle

 \Sec{Introduction.}
\label{sec:intro}
Unveiling the true nature of dark energy (DE) and  dark matter (DM) undoubtedly  
stands as one of the most formidable endeavors in contemporary fundamental 
physics. DM is often explained in terms of elementary particles 
with highly suppressed interactions with the Standard Model (SM) sector.
A particularly compelling  candidate is 
the QCD axion (see \cite{Di_Luzio_2020} for a review), 
a hypothetical particle predicted by the most elegant solution to the strong CP problem~\cite{Callan:1976je,Jackiw:1976pf}, namely the Peccei-Quinn (PQ) mechanism~\cite{Peccei:1977hh,Peccei:1977ur,Weinberg:1977ma,Wilczek:1977pj}.
In addition to offering a clean explanation of the absence of CP violation in strong interactions, axions possess the requisite properties to constitute the entirety of dark matter~\cite{Abbott:1982af,Dine:1982ah,Preskill:1982cy}. 
On another note, DE remains an even deeper mystery (see Ref.~\cite{Huterer:2017buf} for a review). The most economical explanation is a cosmological constant $\Lambda$ 
(see Ref.~\cite{Carroll:2000fy}), which, together with a cold DM (CDM)
component, constitutes the standard  $\Lambda$CDM model.   
This framework has been remarkably successful in accounting for a wide range of cosmological observations.
However,  $\Lambda$CDM faces  significant theoretical challenges concerning the nature of its energy component~\cite{Carroll:2000fy},  and  in recent years its  consistency with observations has  come under increasing  tension  due to high-precision cosmological data.

In this Letter, we propose a scenario in which both DM and DE  are  explained in terms of elementary particles: specifically,  the QCD axion $\varphi_a$ and a dark-sector (DS) $b$-axion $\varphi_b$. This two-axion system exhibits a distinctive and intertwined dynamical behavior.
In recent cosmological times, a  fraction of DM is converted 
into $\varphi_b$ particles through the dynamics of a level crossing (LC), 
thereby populating this new degree of freedom. Once a non-negligible  
energy density $\rho_b$ is generated, it continues to be sourced by 
a creation $C$-tensor, which  is added to Einstein's equation.
We demonstrate that, for suitable values of a new fundamental parameter 
$\eta$, this mechanism can give rise to a transient period of 
$\rho_b$ domination, during which 
$\rho_b$ remains approximately constant and effectively mimics  
a cosmological constant. 
Our mechanism is  inspired by Hoyle's field-theoretic model~\cite{Hoyle:1948} for a steady-state Universe~\cite{Hoyle:1948,Bondi:1948,Hoyle1962,Hoyle1964c}. 
However, it departs from the original proposal in several key aspects. 
First, our construction does not entail a steady-state cosmology: the evolution of the Universe follows  standard Big Bang evolution down to redshifts of $z \sim \text{a few}$, ensuring that all 
the successful features  of the $\Lambda$CDM  model, such 
as CMB generation, structure formation, and other early-Universe processes, 
remain unaltered. Second, our creation $C$-tensor is not constant; rather, it depends  on $\rho_b$ and thus exhibits  space-time dependence.
As a result, $\rho_b$ generation is a local process,  more naturally interpreted  
as the  mass growth of particles that retain a constant number density
(per coming volume), rather than the diffuse creation 
of known particles, such as neutrons or hydrogen atoms, as envisioned in the original proposal~\cite{Hoyle:1948,Bondi:1948}.

In the following, we first present the mechanism of $\rho_b$ generation 
for the DS $b$-axion  
within a modified GR framework and  demonstrate how it can drive accelerated expansion.
Next we describe the dynamics of a $\varphi_a-\varphi_b$ LC, responsible  
for converting a small fraction of CDM into an initial  population 
of $\varphi_b$ particles, which are necessary to trigger the creation mechanism. The particle physics interpretation of the DE phenomenon that we are suggesting
neatly accounts for the DE/DM coincidence puzzle, and it is consistent with  
variations of the effective DE equation of 
state~\cite{DESI:2024mwx,DESI:2025zgx,DESI:2025fii}.
It also predicts that accelerated expansion is a transient phenomenon that does not pose a cosmological problem~\cite{Carroll:2003st}, as the expansion of the Universe will eventually revert to a decelerating phase in the far future.

\Sec{Modified Friedmann Equations.}
We assume a Friedmann–Lema\^itre–Robertson–Walker (FLRW) metric
\begin{equation}
\label{eq:metric}
ds^2 = g_{\mu\nu}dx^\mu dx^\nu =  dt^2 - R^2(t)(dx_1^2 + dx_2^2 + dx_3^2)\,.   
\end{equation}
We  introduce a `creation' vector $C_\mu = (\rho_b,0,0,0)$
where $\rho_b$ represents the energy density of a specific substance. 
From the covariant derivative we define the associated tensor  $C_{\mu\nu}\equiv C_{\mu;\nu} = \frac{\partial C_\mu}{x^\nu} -\Gamma_{\mu\nu}^\alpha C_\alpha$.\footnote{We follow the conventions of Landau and 
Lifshitz~\cite{Landau:1975pou}. 
In particular $\Gamma^\alpha_{\mu\nu} = 
\frac{1}{2}g^{\alpha\sigma}( \partial_\mu g_{\sigma\nu}
+\partial_\nu g_{\sigma\mu}-\partial_\sigma g_{\mu\nu})$.} 
 The only non-vanishing components of 
$C_{\mu\nu}$ are:
\begin{eqnarray}
\label{eq:C}
C_{00} =  {\dot\rho_b}  \,, &\quad & 
C_{ii} = - R \dot R \;{\rho_b}\,,
    \end{eqnarray}
where a dot denotes derivative with respect to time. 
We modify Einstein equation  by adding a $C$-tensor term:
\begin{eqnarray}
\label{eq:fieldeq}
	R_{\mu\nu} -\frac{1}{2} g_{\mu\nu} \mR - \frac{1}{\eta}\,C_{\mu\nu}=  \frac{1}{m_\text{P}^2} T_{\mu\nu}\,, 
\end{eqnarray}
where $R_{\mu\nu}$ denotes the Ricci curvature tensor, $\mR$ 
is the  scalar curvature,
$\eta$ is a new fundamental constant with  dimensions of  number density, 
and $m_P=(8\pi G_N)^{-1/2}$ is the reduced Planck mass.
We are interested in the late matter-dominated and present 
DE-dominated eras, where all forms of 
radiation are subdominant, and their 
contributions to the stress energy tensor $T_{\mu\nu}$ can be neglected. 
Hence, we write $T_{\mu\nu} = 
 T^b_{\mu\nu} +\Tm_{\mu\nu}$,  
 where $T^b_{\mu\nu} =\diag(\rho_b,0,0,0) $ is the stress-energy tensor for the  $b$-substance, assumed to have negligible pressure, while 
$T_{\mu\nu}^m = \diag(\rho_m,0,0,0)$, which we assume to be covariantly conserved, describes all other matter components. 
The covariant derivative of \eq{eq:fieldeq}  
then gives $-\frac{{m_\text{P}^2}}{\eta}(C^{\mu\nu})_{;\nu} =  (T^{b\,\mu\nu})_{;\nu}$, which,  when 
 read right-to-left, indicates 
that whenever $C_{\mu\nu}\neq 0$ (which implies $(C^{\mu\nu})_{;\nu} \neq 0$) 
$b$-substance is created. 
For a FLRW metric, only the components  $(ii)$ and $(00)$ of \eq{eq:fieldeq} are non-vanishing. They are, respectively:
\begin{align}
   2R\ddot R + \dot R^2 - R \dot R\frac{\rho_b}{\eta} = 0\label{eq:ij}\\
  3\frac{\dot R^2}{R^2} = \frac{\rho}{\mP^2} + \frac{\dot \rho_b}{\eta} \,,
  \label{eq:00}
\end{align}
where $\rho=\rho_m + \rho_b$.
Let us rewrite the first equation as:
\begin{equation}
\label{eq:ij2}
2 \frac{d}{d t} \left(\frac{\dot R}{R}\right) + 3 \frac{\dot R^2}{R^2} -\frac{\dot R}{R} \frac{\rho_b}{\eta} =0\,.
\end{equation}
This shows that if at some point during cosmic evolution the condition 
$\rho_b/\eta \approx  3 \dot R/R $ is met, then $\dot R/R$ would be approximately constant, implying an  exponentially accelerated expansion. 
Note that, for consistency, $\rho_b$ must also remain approximately constant during this period.
\eq{eq:00} then shows 
that cosmic evolution can  easily align with such a regime  when the total 
energy density is dominated by $\rho_b$, so that the right-hand (RH) side of this equation also remains approximately constant.
\begin{figure*}[t!]
        \centering
        \includegraphics[width=0.45\linewidth]{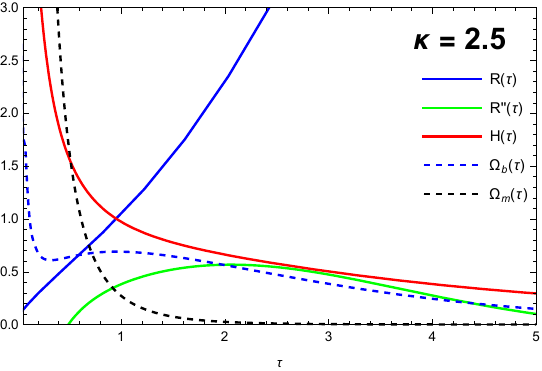}
        \label{fig:plot250}
    \hfill
        \centering
        \includegraphics[width=0.45\linewidth]{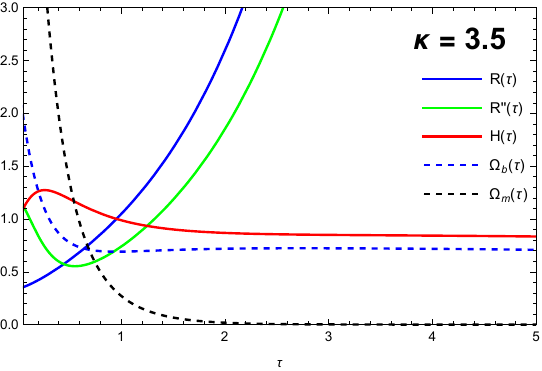}
        \label{fig:plot350}
    \caption{\small Evolution of $R(\tau),\, R''(\tau),\,H(\tau)$ 
    (respectively blue, green and red solid lines) 
    and of the normalized densities $\Omega_b(\tau)$,  $\Omega_m(\tau)$ 
    (respectively blue and black dashed lines)  for the two choices 
    $\kappa = 2.5$ (left plot) and $\kappa = 3.5$ (right plot).
    }
    \label{fig:plots}
\end{figure*}
Let us now see  for which  values of $\eta$ 
the  appropriate condition $\rho_b/\eta \approx  3 \dot R/R $
can be approximately realized at present time $t=t_0$. 
Denote the present value of the Hubble parameter as $H_0 = \frac{\dot R}{R}\big|_{t_0}$ and the present critical density as $\rho_c^0 = 3 H_0^2 \mP^2$.
We can make an educated guess about the relevant range of values for $\eta$ 
by adopting the approximation $ \rho_b(t_0)\approx \rho_c^0$, in which 
$b$-substance saturates the critical energy density.
 We  obtain $\eta \approx  H_0 \mP^2$, which relates  the fundamental  constant $\eta$ to the present value of  
 the Hubble parameter. 
To solve  \eqs{eq:ij}{eq:00} numerically, 
it is convenient to introduce the dimensionless `time' variable 
$\tau = H_0 t$ and a function $\mFb$, which, 
similarly to  $1/R^{3}$ for $\rho_m$, 
describes the evolution of  $\rho_b$: 
\begin{equation}
\label{eq:Omegas}
    \rho_m(\tau) =\rho_c^0\,\Omega_{m,0}  /R(\tau)^{3} \!\! , \quad 
      \rho_b(\tau) =\rho_c^0\, \Omega_{b,0} \,\mFb(\tau)\,,
\end{equation}
where  $\Omega_{m,0}\simeq 0.31$ and $\Omega_{b,0}\simeq 0.69$~\cite{ParticleDataGroup:2024cfk}  are the present fractional energy densities of matter and DE.
Using $H_0 \simeq h (9.8\, {\rm Gyr})^{-1}$, $h\simeq 0.68$, $t_0 \simeq 13.8\,${\rm Gyr}, 
the present value of $\tau$  is 
  $\tau_0  = H_0 t_0  \simeq 0.958$. 
Denoting   derivatives with respect to $\tau$ with a prime, we can  rewrite \eqs{eq:ij}{eq:00} as
\begin{align}
    2RR'' + R'^2 -  \kappa\,RR'\, \Omega_{b,0}\, \mFb= 0\,, \label{eq:H0t3}\\
    \frac{R'^2}{R^2} = \frac{\Omega_{m,0}}{R^3} + 
    \Omega_{b,0}\left(\mFb  +  \frac{\kappa}{3} \mFb' \right)
    \label{eq:H0t4} \,,
\end{align}
where $\kappa = \frac{\rho^0_c}{H_0 \eta}$  is a dimensionless constant, which, for $\eta \approx H_0\mP^2$, takes values around $\kappa \approx 3$. 
The  boundary conditions at   $\tau=\tau_0 $ are given by:
\begin{equation}
 R(\tau_0) = \mFb(\tau_0)= 1;\quad R'(\tau_0)= \frac{dR}{H_0dt}\bigg|_{\tau_0} =  1\,.
 \label{eq:ic}
\end{equation}
The results of  
integrating \eqs{eq:H0t3}{eq:H0t4} with the boundary conditions in  \eq{eq:ic}  are presented in  
Fig.~\ref{fig:plots} for the two representative values 
$\kappa =2.5$ (left panel) and $\kappa =3.5$ (right panel). 
The  plots show the evolution of 
the scale factor $R(\tau)$ (blue line),  the acceleration 
$R''(\tau)$ (green line),  the Hubble parameter $H(\tau)$ (red line), and  
the energy densities $\Omega_b(\tau)=\Omega_{b,0}\mFb(\tau)$  and   $\Omega_m(\tau)= \Omega_{m,0}/R^3$ (blue dashed and black dashed lines, respectively). 
 For both values of $\kappa$,   
 the point of intersection of the 
 two dashed lines (blue and black), where  $\rho_b = \rho_m$, occurs 
 around $\tau_{DE} \simeq 0.72$, corresponding to $\Rde \simeq 0.77$ and 
 $\zde \simeq 0.3$.
 Since  $\Omega_b(\tau)$ remains approximately constant from 
  $\tau_{DE} $  until the present epoch, 
  $\Rde$ 
  can also be estimated straightforwardly as in the 
  $\Lambda \text{CDM}$ model, as  
  $\Rde \approx (\Omega_{m,0}/\Omega_{b,0})^{1/3}$.
For $\tau \gsim  \tau_{DE} $, the rate of decrease of   
 $\rho_b$ (and $H$) in the left panel ($\kappa =2.5$)  is rather mild, 
 powering an accelerated expansion (positive values of the $R''$ green line)
 that lasts until $\tau \sim 6$ (not shown in the plot).  
For $\tau>6$ the expansion transitions to deceleration.  
Thus acceleration is a transient phenomenon 
that begins around $\tau \sim 0.5$, while at earlier times $\tau\lesssim 0.5$,  the expansion was decelerating.  

For $\kappa =3.5$, both  $\rho_b$ and $H$\ remain 
remarkably constant from  $\tau_{DE}$  until $\tau \gsim 10$. 
In the regime where $\rho_b\gg \rho_m$, a constant value of $\rho_b$ powers an exponential expansion, 
as can be verified  by substituting $R(\tau)\approx \exp(\frac{\rho_b}{3 H_0\eta} \tau)$ into \eq{eq:H0t3}, or by observing   that in the 
RH panel in Fig.~\ref{fig:plots}, the $R''$ green line 
 closely tracks the $R$ blue line. 
In this regime, the role of the $C$-tensor in driving cosmic 
acceleration becomes particularly clear. 
For $\tau \gsim \tau_0$ we have $\dot \rho_b \approx 0$, and  
since the contribution 
of $\rho_m$ is strongly suppressed,  $\rho \approx \rho_b$. 
We can then move the $C$-tensor to the 
RH side of \eq{eq:fieldeq} and define:
\begin{equation}
    \widetilde{T}^b_{\mu\nu} = T^b_{\mu\nu} + \frac{\mP^2}{\eta} C_{\mu\nu}\,. 
\end{equation}
From the second relation in \eq{eq:C}  we obtain 
\begin{equation}
 C_{ii} = - \;R^2\, \frac{\dot R}{R}\rho_b \approx - R^2\; 3 \eta \frac{\dot R^2}{R^2}
\approx - \frac{\eta}{\mP^2} \;R^2\, \rho\,, 
\end{equation}
where we have used $\rho_b/\eta \approx  3 \dot R/R $  and   \eq{eq:00}
in the last step.   Thus, to the extent that  the approximations 
 $\dot \rho_b \approx 0$ and  $\rho_b \approx  \rho$ hold, 
 we can write    $\widetilde{T}^b_{\mu\nu} \sim \rho_b \;g_{\mu\nu}$, 
 which shows that in this regime,  
$\widetilde{T}^b$ mimics a positive cosmological 
constant.
For $\kappa=3.5$, cosmological acceleration is also 
a transient phenomenon, with the expansion transitioning to deceleration  
around $\tau \lesssim 20$  (not shown in the plot).
Thus,  in the range  $2.5\lesssim \kappa\lesssim 3.5$ the solutions remain  well behaved in the limit $\tau \to \infty$, and predict that all forms of 
energy density will eventually be diluted away. 

In order to  identify some peculiar properties of the $b$-substance,  
it is helpful to study the early times initial conditions. We find that for both values of $\kappa$, the ratio $\Omega_b/\Omega_m$ remains remarkably constant as long as $\tau \lesssim 0.1$, corresponding  to a 
  redshift  $z\simeq 4$ for $\kappa =2.5$,   
  and $z\simeq 2$ for $\kappa=3.5$. 
  Thus,  at early times, the behavior of  the $b$-substance closely tracks 
  that of matter, i.e. $\rho_b \sim R^{-3}$.
Moreover,  for  $\kappa =2.5$ we find  $(\Omega_b/\Omega_m)_{\tau<0.1} \sim 2\%$,  while 
for  $\kappa =3.5$  $(\Omega_b/\Omega_m)_{\tau<0.1} \sim 28\%$. 
That is, the initial abundance of the 
$b$-substance required by present-day boundary conditions is a non-negligible fraction of the matter density.
In the next section, we  describe a mechanism that can naturally generate 
a density of DS particles with this characteristic.

\Sec{Mechanism for generating the initial conditions.}
\label{sec:twoaxions}
The cosmic acceleration mechanism described above requires 
a certain initial amount of $b$-substance, ranging from a few percent to a few tens of percent of the matter energy density,  to ensure that,  in~\eq{eq:fieldeq}, a non-vanishing $C$-tensor  can drive the creation of  $\rho_b$ and match the boundary conditions set at the present epoch. 
Moreover, the initial $b$-substance seed must emerge at a relatively recent cosmological time, $t \lesssim \tde$, since 
a precocious appearance at $t \ll \tde$ would result in a distinct and cosmologically inconsistent early phase of accelerated expansion.
These essential requirements can be  satisfied by identifying the $b$-substance with an axion $\varphi_b$ coupled to a strongly interacting 
gauge group, $G_b$, of a dark 
sector that underwent confinement in a recent epoch.\footnote{With an appropriate fermion spectrum, the coupling constant of a dark gauge group of small order, like e.g. $SU(2)$, can decrease towards small values until some of the fermions are integrated out, naturally delaying confinement to a scale below the eV.} 
With such a low confinement scale, the axion misalignment mechanism would be 
unable to produce any substantial amount of $b$-axions. However, a   
 useful  $\varphi_b$  number density 
 can be generated via the dynamics of a  non-adiabatic 
LC with the QCD axion, which is assumed to comprise the entirety of 
DM.\footnote{Two axion LC, typically  
in the  adiabatic regime,  has been previously harnessed in 
various contexts unrelated to DE~\cite{HILL1988253,Kitajima:2014xla,Daido:2015bva,Daido:2015cba,
Higaki:2016yqk,
Ho:2018qur,Cyncynates:2021xzw,Cyncynates:2023esj,Li:2023uvt}.}    
We now outline the main features of a simple two-axion model that realizes this scenario; a more general and detailed treatment is provided in the Appendix.

Let us denote $G_a = SU(3)_{\rm QCD}$ and   
introduce a  dark gauge group $G_b=SU(2)$,  
with confining scale $\Lambda_{b} \ll \Lambda_{a}$.   
Consider the following  Yukawa Lagrangian 
\beq
\label{eq:Lyuka}
\mathcal {L}_Y = \bar \psi_L \psi_R \Phi_1 + \bar\chi_L \chi_R \Phi_2\, ,
\eeq
where the  fermions 
transform  under $G_a \otimes G_b$ as 
$\psi_{L,R} \sim  (\mathbf{1}, \mathbf{2})$ and 
 $\chi_{L,R} \sim (\mathbf{3}, \mathbf{2})$, while    
$\Phi_{1,2}$ are two gauge singlet scalars acquiring VEVs $\varv_{1,2}$. 
The Lagrangian enjoys two Peccei-Quinn (PQ) symmetries 
$U(1)_q\otimes U(1)_p$ under which the fields transform with global charges 
$\Phi_1 \sim (q_1, 0)$, $\Phi_2 \sim (0, p_2)$  and 
$q_{\psi_L} - q_{\psi_R} = q_1$,
$p_{\chi_L} - p_{\chi_R} = p_2$. Without loss of generality
we can normalize $q_1=p_2=1$.
Since $\psi_{L,R}$ do not carry color,  there is no  
$U(1)_q$--\,$G_a$ mixed anomaly. The coefficients of the 
other   anomalies 
$U(1)_q$-\,$G_b$, $U(1)_p$-\,$G_a$ and $U(1)_p$-\,$G_b$ are 
respectively
\begin{equation}
n_{qb}   
= 1\,, \quad
n_{pa}  
= 2\,,   
\quad
n_{pb}   
= 3\,. 
\end{equation}
After  $U(1)_{q,p}$ spontaneous breaking,   the Yukawa terms in \eq{eq:Lyuka}
give rise to the effective Lagrangian
\beq
\label{eq:LyukEff}
\mathcal {L}_Y^{\rm eff} = \bar \psi_L \psi_R \varv_1 e^{i
\frac{a_1}{\varv_1}} + \bar\chi_L \chi_R \varv_2 e^{i 
\frac{a_2}{\varv_2}}\, .
\eeq
Removing the phases via chiral rotations generates the anomalous terms 
$ \frac{\mathcal{C}_{i}}{16\pi^2}\,  F_i\cdot \tilde F_i $ ($i=a,b)$
with coefficients:
\begin{equation}
\label{eq:a-anomaly}
 \mathcal{C}_a  = 
 n_{pa}\frac{a_2}{\varv_2}, \qquad
 \mathcal{C}_b  = n_{qb}\frac{a_1}{\varv_1} +  n_{pb} \frac{a_2}{\varv_2} \,. 
\end{equation}
Since, by assumption,  $\Lambda_a \gg \Lambda_b$,  at a temperature $T \sim \Lambda_a$ the field  $\varphi_a \equiv a_2$ acquires a mass from non-perturbative QCD effects,
while $\varphi_b \equiv a_1$ remains effectively massless.
 Let us rewrite 
\beq
\label{eq:Cab}
\mathcal{C}_a = \frac{\varphi_a}{F},\qquad \mathcal{C}_b = \frac{\varphi_a}{F'} + 
\frac{\varphi_b}{f}  \,,
\eeq
where 
\beq
\label{eq:FFf}
F=\frac{\varv_2}{2}, \qquad
F'=\frac{\varv_2}{3}, \qquad
f=\varv_1\,.
\eeq
As the temperature  decreases toward  $\Lambda_b$, the
effects of $G_b$ instantons become relevant,   
 and a potential is generated  for $\varphi_b$ as well.  
The full potential can be written as:
\beq
\label{eq:potentialN}
 \hspace{-0.2cm} 
V = \Lambda^4_a  \left[1-\cos \left(\frac {\varphi_a}{F} \right)\right] +
\Lambda^4_b \left[1-\cos \left( \frac {\varphi_a}{F'}+\frac {\varphi_b}{f} \right)\right]. 
\eeq
From $V$   one can derive the equations of motion that, 
in the limit of small oscillations, can be written as:
\beq
\label{eq:oscillatorsN}
\ddot A + 3 H \dot A + \mM^2 A =0\,,
\eeq
with   
\beqa 
\label{eq:massmatrixN}
A=\begin{pmatrix}
\varphi_a \cr \varphi_b
\end{pmatrix} \,, \qquad
\mM^2 = 
m^2_a 
\begin{pmatrix}
1 +  \epsilon^2  r  & 
 \epsilon\, r
\cr 
 \epsilon\, r & r
\end{pmatrix}\,.
\eeqa
We have defined $m_a =\Lambda^2_a/F$,   $\epsilon = f/F'$ 
and $r=r(T)=m^2_b(T)/m^2_a$, where we have kept the temperature 
dependence of $m_b$.
We now assume  that at zero temperature  $m_b = \Lambda^2_b/f > m_a$. 
Since $\Lambda_a \gg \Lambda_b$,    it follows that   $f \ll F$, which  
is easily realized by assuming 
$ \varv_1\ll \varv_2$, see \eq{eq:FFf}.  
Clearly, at a certain temperature  $\Lambda_a \gg \Tlc \gtrsim \Lambda_b$, when  $m_b(T)$ is still evolving, the condition $m_b(\Tlc)=m_a$ is reached. 
Then, neglecting the highly suppressed  $\epsilon^2$ term,  we have 
$\mM^2_{11}=\mM^2_{22}$, which triggers a LC.

 Let us  now introduce the time variable $x=t/\tlc$. 
We are interested in a LC occurring during matter domination
($R(t)\sim t^{2/3}$).  
Let us
 consider the evolution around $x \sim 1$ and  write 
   $m^2_b(T) = m^2_a (\Tlc/T)^n$,   so that
 $r(x) =x^{\frac{2n}{3}}$. 
 The exponent $n$ depends on the 
   gauge group and the particle content of the model,  typically ranging between 4 and 8. Since the precise value of the exponent is not crucial to implement the LC mechanism,     
  we take $n=6$ as suggested by 
  QCD lattice simulations  for temperatures around the onset 
  of  axion oscillations~\cite{Petreczky:2016vrs,Berkowitz:2015aua}.
   \eq{eq:oscillatorsN} can now be written as: 
 \begin{align}
 \label{eq:oscillators2a}
 & \hspace{-1cm} \ddot  \varphi_a + \frac{2}{x} \dot \varphi_a + \omega^2 \varphi_a + \epsilon \omega^2 x^4
 \varphi_b = 0,\\ 
  \label{eq:oscillators2b}
&   \hspace{-1cm} \ddot  \varphi_b + \frac{2}{x} \dot \varphi_b + \omega^2 x^4 \varphi_b + \epsilon \omega^2 x^4 \varphi_a = 0,
\end{align}
where $\omega = m_a \tlc$ and the dots represent derivatives with respect to $x$.
In the region around LC, where $m_b(\Tlc)\simeq m_a$, 
 the splitting in the mass eigenvalues is determined by 
the $O(\epsilon)$ off-diagonal entries in the mass matrix. 
As discussed in more detail in the Appendix,  if the splitting is much larger than 
the  variation of the mass ratio $r(T)$ in the resonance region, the heavier state at $t \ll \tlc$
($\varphi_a $) remains the heavier, and emerges as $\varphi_b$ at $t \gg \tlc$. That is,  in crossing the resonance region,
the two axions swap their ``flavor'' identities. 
This defines the adiabatic regime, for 
which  $ \epsilon\, \omega  = \epsilon\,\tlc\, m_a \gg 1$.
Since $\epsilon\, \tlc $ is the  width of the resonance, and $m_a$ the  
oscillation frequency at LC, 
adiabaticity requires that  several oscillations occur within the resonant region.  
Conversely, if  $ \epsilon\, \omega$ is small, typically not much larger than order one, the transition is non-adiabatic, flavor conversion is not efficient, and only a fraction of $\varphi_a $ is  converted into $\varphi_b$. 
In our scenario, stringent conditions ensure that the LC proceeds in 
the non-adiabatic regime. 
The evolution of $m_b(T)$ around $\Tlc$ can be written as:
\begin{equation}
m_b(\Tlc) \sim   \frac{\Lambda_b^2}{f} \left(\frac{T_b}{\Tlc}\right)^{3} = m_a = \frac{\Lambda_a^2}{F} \,.
\end{equation}
where $T_b\approx \Lambda_b$ is the  $G_b$ confinement temperature. 
We take for simplicity the same temperature for the $G_a$ and $G_b$ sectors (we will comment later on this point). Let us take 
 $F' \sim F \sim v_2$ and  $f = v_1$ and  
assume   that $G_b$ confines not long after LC: $(T_b/\Tlc)^3\sim 10^{-1}$.
We have 
\begin{equation}
\label{eq:epsilonN}
\epsilon = \frac{f}{F'}\sim 0.1 \frac{\Lambda_b^2}{\Lambda_a^2} 
\sim  10^{-24} \left(\frac{\Lambda_b}{10^{-3}\mathrm{eV}}
\,\frac{200\mathrm{MeV}}{\Lambda_a}\right)^{2},
\end{equation}
 An  additional requirement is 
that  the PQ $U(1)_b$ symmetry is spontaneously broken at a temperature sufficiently above  $\Tlc$. Since $\Tlc > \Tde \sim 3
\cdot 10^{-4}\,$eV,  
let us take $f\gsim 5\cdot 10^{-3}\,$eV.  
Together with \eq{eq:epsilonN}, this yields 
$F\sim 5\cdot 10^{12}\,$GeV, which  corresponds to $m_a \sim  10^{-6}\,$eV
that is in the ballpark for reproducing 
$\Omega_{DM}$ from the misalignment mechanism 
in the post-inflationary   QCD axion scenario. Larger values of   $f$ can still be consistent with axion DM,  assuming a  pre-inflationary 
scenario with a moderate tuning 
of the initial misalignment angle. 
Using \eq{eq:epsilonN}  and taking  values $\taulc \lesssim 0.1$
($\zlc\gsim 2$) 
we obtain 
$ \epsilon\, \omega  \sim O(1 - 10)$. 
Hence, the LC  proceeds in the non-adiabatic regime, and only 
a fraction of QCD axions is converted into 
 $\varphi_b$, providing  
 initial conditions consistent with 
the requirements of the previous section.

\Sec{Discussion.}
To explain cosmic  acceleration, we have introduced 
a dark-sector $b$-axion whose energy density is continuously generated 
through an increasing mass, an effect that is implemented 
via the addition of a creation $C$-tensor to Einstein’s equations.
While axions coupled to a strongly interacting gauge group via an anomalous 
term are known to undergo mass-growth near the confinement 
phase transition, we exploit this feature solely   
to generate an initial number density $n_b$, required to 
trigger the creation mechanism. This is achieved through 
a non-adiabatic LC with the QCD axion taking place at a relatively recent 
cosmological epoch, during which a small fraction of DM is 
converted  into $\varphi_b$. Before the LC, the  cosmological evolution 
 follows the standard history, apart from a suppressed 
extra radiation contribution, compatible with current 
limits on the effective number of neutrino 
species~\cite{DESI:2024mwx}.\footnote{The SM and the DS  
interact via QCD-dark gluon scattering  $g_ag_a \leftrightarrow g_bg_b$
 induced by  loop box diagrams involving the $\chi$ fermions.  
Since   $m_\chi\sim v_2 \gsim 10^{12}\,$GeV, 
the two sectors get thermally decoupled well before 
the annihilation of $O(100)$ SM degrees of freedom reheats the visible sector,  implying $T_{\rm DS} \ll T_0$.}
After the LC, the total QCD axion DM abundance is slightly reduced. However,  
given the considerable uncertainties in determining  the present-day DM 
content of the Universe,  this effect is unlikely to be distinguishable. 
As for the cosmic distribution of $b$-axions, we should not expect 
 it to follow that of DM. This is because their decay constant 
$f$ is not hierarchically  larger than their mass $m_b$, implying that 
 their   self-coupling $\lambda m_b/f$ is not strongly suppressed. As a result, self-interactions (as well as interactions with  other DS particles) may 
efficiently transfer energy to non-zero momentum modes~\cite{Abbott:1982af},
causing $b$-axions to diffuse out of the gravitational potential wells 
where DM clusters. We leave this important aspect for future work.
Besides offering a novel interpretation  of the DE phenomenon, 
our framework also yields some encouraging implications.
Notably, it predicts  DE time-variance  at a late epoch, 
a behavior that  recent results from the DE Spectroscopic Instrument (DESI) 
collaboration suggest may be favored over the standard $\Lambda$CDM 
model~\cite{DESI:2024mwx,DESI:2025zgx,DESI:2025fii}.
In particular, the evolution of the normalized energy density $\rho_{\rm DE}(z)/\rho_{\rm DE,0}$ found in Ref.~\cite{DESI:2025fii} - namely the increase with the scale factor, the broad peak around 
$\zde$, and the decrease as the Universe continues to expand - is qualitatively reproduced by 
our model with $\kappa = 2.5$ (see, however, Ref.~\cite{Nesseris:2025lke}).
This model can also account for a recently reported  $\sim 5\sigma$ tension.
Using cosmological datasets,  Ref.~\cite{Mukherjee:2025fkf} reconstructed 
the Universe's expansion rate at two different redshifts:
  $z_1 = 1.646$ (where the angular diameter distance $D_A$ reaches its maximum) 
and  $z_2 = 0.512$  (where $d D_A/d z =D_A$). 
 In the $\Lambda$CDM model,  the predicted  value of $H(z_2)$  
 deviates from  the reconstructed one at the level  of $5\sigma$.
 In contrast, we find that without any tuning of the parameters,  
for  $\kappa = 2.5$, the predicted values of both  $H(z_1)$ and $H(z_2)$
lie within $1\sigma$ of the reconstructed values.


\section*{Acknowledgments}
The work of E.N.~was supported  by the 
Estonian Research Council grant PRG1884   
and by the INFN ``Iniziativa Specifica" Theoretical Astroparticle Physics (TAsP-LNF). The work of K.M.~was supported  by the 
Estonian Research Council grant PRG803 and  the Estonian Research Council personal grant PUTJD1256. We acknowledge support from the CoE grant TK202 “Foundations of the Universe” and from the 
CERN and ESA Science Consortium of Estonia, grants RVTT3 and RVTT7.   
This article is based in part upon work from COST Action COSMIC WISPers CA21106, supported 
by COST (European Cooperation in Science and Technology).
We thank the Galileo Galilei Institute for Theoretical Physics, where this work was started, for hospitality.


\input{Appendix.tex}


 \bibliographystyle{apsrev4-1}

 \input{Letter_PRLv3.bbl}

\end{document}

%% file: Appendix.tex




\bigskip
\bigskip

 \centerline{\bf APPENDIX}

\bigskip
\smallskip

Here we discuss a generalization of the two axion  
construction discussed in the main text, and we study in 
more depth the dynamics of the LC.  We assume 
two generic confining groups $G_{a,b}$ 
 and  two set of fermions 
transforming   in generic $G_a \otimes G_b$ 
 representations  
$\psi_{L,R} \sim [(d^a_\psi,\,T^a_\psi),\, (d^b_\psi,\,T^b_\psi)]$,  
 $\chi_{L,R} \sim [(d^a_\chi,\,T^a_\chi),\, (d^b_\chi,\,T^b_\chi)]$,
where   $d^{a,b}$ and $T^{a,b}$ denote respectively the dimension 
and index of the representation. The  Yukawa Lagrangian 
\eq{eq:Lyuka} repeated here for convenience 
\beq
\label{eq:Lyuk}
\mathcal {L}_Y = \bar \psi_L \psi_R \Phi_1 + \bar\chi_L \chi_R \Phi_2\, ,
\eeq
involves in total six fields carrying six overall phases.
Two conditions for  rephasing  invariance are fixed    
by~\eq{eq:Lyuk}. The remaining 
four global symmetries are  two independent baryon numbers $B_\psi,\,B_\chi$, and  
two global Peccei-Quinn (PQ) symmetries 
$U(1)_{q,p}$ under which the scalar fields carry charges 
$\Phi_1 \sim (q_1, 0)$ and $\Phi_2 \sim (0, p_2)$.
The Yukawa terms impose the conditions 
$q_{\psi_L} - q_{\psi_R} = q_1$,
$p_{\chi_L} - p_{\chi_R} = p_2$ and, without loss of generality, 
we can normalize $q_1=p_2=1$. 
Then the mixed $U(1)_{q,p}-G_{a,b}$ anomaly coefficients are given by:
\begin{equation}
\!\!\! n_1 \!= 2 d^b_{\psi}  T^a_\psi, 
\
m_1 \!= 2 d^a_{\psi} T^b_\psi, 
\
n_2 \!= 2 d^b_{\chi}  T^a_\chi,  
\
m_2 \!= 2 d^a_{\chi}  T^b_\chi,
\end{equation}
where the factors of 2 ensure that 
the coefficients are integers.
After  $U(1)_{q,p}$ spontaneous breaking   the Yukawa terms in \eq{eq:Lyuk}
give rise to the effective Lagrangian
\beq
\label{eq:LyukEffApp}
\mathcal {L}_Y^{\rm eff} = \bar \psi_L \psi_R \varv_1 e^{i
\frac{a_1}{\varv_1}} + \bar\chi_L \chi_R \varv_2 e^{i 
\frac{a_2}{\varv_2}}\, .
\eeq
Removing the 
phases via chiral rotations 
generates the anomalous terms 
$ \frac{\mathcal{C}_{i}}{16\pi^2}\,  F_i\cdot \tilde F_i $ ($i=a,b)$
with coefficients:
\begin{equation}
\label{eq:a-anomalyApp}
 \mathcal{C}_a  = 
n_1 \frac{a_1}{\varv_1} + n_2\frac{a_2}{\varv_2}, \qquad
 \mathcal{C}_b  = 
m_1\frac{a_1}{\varv_1} +  m_2 \frac{a_2}{\varv_2} \,. 
\end{equation}
%
%
%
\begin{table}[t!!]
\begin{center}
\begin{tabular}{ |l|l| } 
\hline  
$ \phantom{\Big|}SU(2):$ 
& $ {\bf 2}\,(1)\ \ \ {\bf 3}\,(4)\ \ \  {\bf 4}\,(10)\ {\bf 5}\,(20)\ \ {\bf 6}\,(35)\ \ \ \  {\bf 7}\,(56) $  \\ [0.3ex]  
$ \phantom{\Big|}SU(3):$  
& $ {\bf 3}\,(1)\ \ \ {\bf6}\,(5)\ \ \  {\bf 8}\,(6)\ {\bf 10}\,(15)\ {\bf 15}\,(20)\  {\bf 21}\,(35) $  \\ [0.3ex]  
$ \phantom{\Big|}SU(4):$  
&  $ {\bf 4}\,(1)\ \ \  {\bf 6}\,(2)\ {\bf 10}\,(6)\   {\bf 15}\,(8)\ \ \  {\bf 20}\,(13)\ {\bf 20'}\,(16) $ \\   [0.3ex] 
$ \phantom{\Big|}SU(5):$  
&  $ {\bf 5}\,(1)\ {\bf 10}\,(3)\ {\bf 15}\,(7)\ {\bf 24}\,(10)\ {\bf 35}\,(28)\ {\bf 40}\,(22) $ \\   [0.3ex] 
 \hline 
\end{tabular}
\end{center}
    \caption{ 
  $SU(N)$ ($N=2,3,4,5$) representations 
of lowest dimension (in bold face) with twice the value of the index given in parenthesis. 
    \label{tab:dimindex}}
\end{table}
We see that if the dimension/index of 
the representations satisfy the 
condition  $(d^a_\psi d^b_\chi)/(d^b_\psi d^a_\chi)=(T^a_\psi T^b_\chi)/(T^b_\psi T^a_\chi)$
(i.e.  ${n_1}/{n_2}= {m_1}/{m_2}$)
then  $ \mathcal{C}_a \propto  \mathcal{C}_b$. The  field combination  orthogonal to $ \mathcal{C}_{a,b}$ 
would then decouple from the symmetry breaking effects generated by the anomalies, maintaining 
a flat potential. 
If instead the relation holds  only approximately, then the flat direction 
 would get lifted only slightly, 
 realising the Kim-Nilles-Peloso (KNP) mechanism~\cite{Kim:2004rp}.  
Clearly, achieving approximate proportionality to a specified degree requires a careful choice of 
groups and representations, as can be understood from the 
dimension and index values of $SU(N)$  representations listed in Table~\ref{tab:dimindex}. 


%
Assuming  $\Lambda_a \gg \Lambda_b$ implies that 
at a temperature $T \sim \Lambda_a$ the field combination 
$\varphi_a/F \sim \mathcal{C}_a$ 
acquires a mass while the orthogonal 
combination $\varphi_b/f\sim - n_2 a_1/\varv_2 + n_1 a_2/\varv_1$ remains massless. 
The  mass eigenstates  $(\varphi_a,\varphi_b)$ are   related to $(a_1,a_2)$ 
by an orthogonal transformation characterized by the angle $ \vartheta = \arctan(\frac{n_2 v_1}{n_1 v_2})$.
Following Ref.~\cite{Choi:2014rja}, we can obtain the respective 
decay constants from the variations of the fields
 $\delta a_{i} = v_{i} \alpha_{i}$ ($i=1,2$):
\begin{equation}
\delta \varphi_a \equiv F \eta = c_\vartheta \delta a_1 + s_\vartheta \delta a_2 =\frac{v_1v_2}{ \sq}
(n_1 \alpha_1 + n_2 \alpha_2)
\end{equation}
where $c_\vartheta (s_\vartheta) =  \cos\vartheta (\sin\vartheta)$, 
 $\sq= \sqrt{n_1^2 \varv_2^2 +n_2^2 \varv_1^2}$,  
and $\eta = n_1 \alpha_1 + n_2 \alpha_2\neq 0$ is an arbitrary non-vanishing shift. Therefore  
$F=v_1 v_2/\sq $.
A variation in the orthogonal direction is identified by the condition 
$n_1 \hat\alpha_1 + n_2 \hat\alpha_2=0$,
that is $\hat\alpha_1 = - n_2 \eta',\ \hat\alpha_2 =  n_1 \eta'$ for some arbitrary shift~$\eta'$: 
%
\begin{equation}
\delta \varphi_b  
= -s_\vartheta \delta \hat a_1 + c_\vartheta \delta \hat a_2 =   \sq \cdot \eta' \,. 
\end{equation}
Finally, by expressing $a_{1,2}$ in $\mathcal{C}_b$ in  \eq{eq:a-anomalyApp}
in terms of $\varphi_a, \varphi_b$ one can easily obtain 
\beq
\label{eq:Cb}
\mathcal{C}_b = \frac{\varphi_a}{F'} + 
\frac{\varphi_b}{f}  \,,
\eeq
with 
\beq
\label{eq:Fp}
 F' = \frac{\varv_1 \varv_2 \; \sq 
 }{m_1 n_1 \varv_2^2 + m_2 n_2 \varv_1^2}, \quad 
 f  = \frac{ \sq 
 }{|m_2 n_1 - m_1 n_2|} \,.
\eeq

\Sec{Equations of motion.} For convenience, we repeat below the expression for potential generated by the $G_a\times G_b$ strong dynamics and the equations of motion for the two axion system 
given in \eqs{eq:potentialN}{eq:massmatrixN} in the main text:
\beq
\label{eq:potential}
\hspace{-0.2cm} V = \Lambda^4_a  \left[1-\cos \left(\frac {\varphi_a}{F} \right)\right] +
\Lambda^4_b \left[1-\cos \left( \frac {\varphi_a}{F'}+\frac {\varphi_b}{f} \right)\right], 
\eeq
\beq
\label{eq:oscillators}
\ddot A + 3 H \dot A + \mM^2 A =0\,,
\eeq
where  
\beqa 
\label{eq:massmatrix}
A=\begin{pmatrix}
\varphi_a \cr \varphi_b
\end{pmatrix} \,, \qquad
\mM^2 = 
m^2_a 
\begin{pmatrix}
1 +  \epsilon^2  r  & 
 \epsilon\, r
\cr 
 \epsilon\, r & r
\end{pmatrix}\,.
\eeqa
Since by assumption $\Lambda_a \gg \Lambda_b$,
requiring that at zero temperature  $m_b = \Lambda^2_b/f > m_a$  --
which is a mandatory condition to ensure a LC -- 
implies  $f/F\ll 1$.
%
A class of models satisfying this condition is easily obtained 
by assuming $v_1/v_2 \ll 1$ and by choosing $\psi$ to be a singlet of $G_a$. Then 
 $n_1=0$ and  $f/F = (n_2/m_1)(v_1/v_2)$, while   
 $\epsilon = f/F' = (m_2/m_1)(v_1/v_2)\ll 1 $.  
It is the clear that at temperatures  $\Lambda_a \gg T \gtrsim \Lambda_b$, when 
$m_a$ has long reached its constant value while 
 $m_b(T)$ is still evolving,  a LC  
will occur. Neglecting for simplicity 
the highly suppressed  $\epsilon^2$ term in $(\mathcal{M}^2)_{11}$, the LC is defined  
by the condition $m_b(\Tlc)=m_a$.
%

 Let us now define the instantaneous mass basis $A_m$ as
 \beq
A= R(x)\; A_m, \qquad  M^2 = R^\dagger(x) \mathcal{M}^2 R(x)\,,
 \eeq
where $M^2 =  \diag (M^2_+, M^2_-)$ and $R(x)$ is an orthogonal matrix defined in terms of an angle $\beta(r(x))$.  
We have:
\beq
M^2_\pm = \frac{m^2_a }{2} (1+r \pm\Delta), \quad \tan\beta = \frac{2 \epsilon r}{1-r +\Delta}
\eeq
with $\Delta = \sqrt{(1-r)^2+ 4 \epsilon^2 r^2}$. At LC  $\tan\beta \to 1$ and 
$R(x=1)$ describes maximal mixing. At $r = (1+\frac{3}{2}\epsilon )^{-1}$ 
$\tan\beta =\frac{1}{2}$, hence the  width of the resonance is $\Delta r\sim 3 \epsilon$. 
 In the  instantaneous mass basis  \eq{eq:massmatrix} becomes
\beq
\label{eq:massoscillators}
 \ddot A_m +  3\mathsf{H} \dot A_m +  \mathsf{M}^2 \!A_m=0\,, 
\eeq
where 
\beq 
 \mathsf{H} = H + \frac{2}{3} R^\dagger\dot R, \qquad  \mathsf{M}^2 = M^2 + R^\dagger \ddot R + 3H R^\dagger \dot R\,. 
\nonumber
\label{eq:HM}
\eeq
and 
\beqa
\nonumber
R^\dagger \dot R &=& i\sigma_2 \cdot \frac{\epsilon}{\Delta^2} \dot r \,, \\
\nonumber
R^\dagger \ddot R &=& i\sigma_2 \cdot \left[\frac{2(1-r-4r\epsilon^2)}{\Delta^2}  
\dot r^2  +\ddot r\right]\frac{\epsilon}{\Delta^2} -\sigma_0\cdot \frac{\epsilon^2}{\Delta^4} \dot r^2 \qquad
\eeqa
with $\sigma_2 $ the  Pauli matrix and $\sigma_0$ the identity. 
The nature of the LC  is characterized by the ratio  between the 
splitting of the two levels and the off-diagonal entries in  $\mathsf{M}^2$ 
that mix these levels, evaluated at LC: 
\beq
\nonumber
\label{eq:LZparameter}
\!\!\!\!  \gamma=
\left| \frac{\tr ( \sigma_3  \mathsf{M}^2)}{\tr (\frac{i}{2}\sigma_2  \mathsf{M}^2)}\right|_{\rm LC} \!\!\! =
\left|\frac{8 m^2_a \epsilon^2}{3 H \dot r - 2 \dot r^2 + \ddot r}\right|_{\rm LC}\!\!\!\!  =  
\frac{36   (\epsilon \tlc m_a)^2}{n (2n-3)}\,.
\eeq
 If the splitting is much larger than 
the  variation of $r$ in the resonance region ($\gamma\gg 1$), the heavier state at $t \ll \tlc$
($\varphi_a $) remains the heavier, and emerges at $t \gg \tlc$ as $\varphi_b$. That is,  in crossing the resonance region,
the two axions swap their ``flavor'' identities. 
This defines the adiabatic regime, for which  $ \epsilon\, \tlc  m_a\gg 1$.
Since $\epsilon\, \tlc $ is the  width of the resonance and $m_a$ the  oscillation frequency at LC, 
adiabaticity requires that  several oscillations occur within the resonant region.  
The adiabatic LC phenomenon is well known in condensed matter 
physics, and analytic treatments valid under certain assumptions 
were formulated long ago, most notably by Landau~\cite{Landau:1932vnv} 
and Zener~\cite{Zener:1932ws}. In particle physics, the phenomenon is realised in the  
Mikheyev-Smirnov-Wolfenstein  (MSW) enhancement of in-matter $\nu_e\to \nu_\mu$ 
conversion of solar neutrinos~\cite{Wolfenstein:1977ue,Mikheyev:1985zog}. 
However, our two-axion LC  differs from the MSW effect in that besides   
the mass splitting 
also   the off diagonal entries in $\mathcal{M}^2$  
are time-dependent. It bears a closer resemblance to a variant MSW realization~\cite{Roulet:1991sm} 
where the off-diagonal entries stem 
 from (hypothetical) $\nu_e\!-\!\nu_\mu$ flavor changing interactions with 
 electrons and nucleons,  and  also  vary as a function of the  matter  density.

\begin{figure}[t!]
\centering
\includegraphics[width=0.45\textwidth]{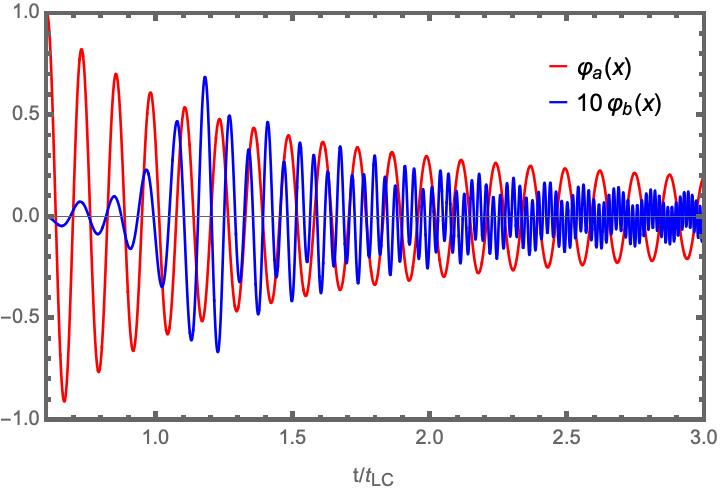}
\caption{An example of the LC mechanism for $\epsilon\, \omega=1$ and $\omega=50$. 
The  $\varphi_b$ amplitude is multiplied by a factor of 10. 
\label{fig:LC}}
\end{figure}

As we have stressed in the main text, consistency conditions 
constrain the LC to occur in the non-adiabatic regime. 
In our scenario, 
a precise study of the dynamics of the  LC in this regime 
is an extremely difficult task. Analytic tools as the Landau-Zener (LZ) 
approximation~\cite{Landau:1932vnv,Zener:1932ws} cannot be employed  
because they rely on assumptions that in our case are not respected  (a mass splitting linear 
function of time and time independent off-diagonal mixing terms). 
Moreover, it is known that the  LZ formula cannot be extrapolated to  the strong non-adiabatic regime~\cite{Rosen:1986jy,Kim:1986wg,Kim:1987ss}. 
On the other hand, attempts at numerical integration of 
the  equation of motion for a two-axion system, such as   
Eqs.~(\ref{eq:oscillators2a})-(\ref{eq:oscillators2b}) 
encounter daunting  obstacles due to the well-known difficulty 
of tracking rapidly oscillating fields at late cosmological times. 
In our case,  this difficulty arises from  the extremely large value of the dimensionless parameter
$\omega$ that, for  the values of $m_a$ and $\tlc$ mentioned above, reaches   
$\omega \sim 10^{24}$. 
An example of a non-adiabtic LC obtained by integrating 
\eqs{eq:oscillators2a}{eq:oscillators2b} 
with the adiabatic parameter fixed at  $\epsilon \omega=1$ is shown in Fig.~\ref{fig:LC}. The red line depicts the evolution   of  the QCD axion amplitude $\varphi_a$ with  boundary condition $\varphi_a(0)=1$. 
The blue line  depicts the evolution of the amplitude 
$\varphi_b$ multiplied by a factor of ten. The initial value  
$\varphi_b(0)$ is  taken to be arbitrarily small.
After LC, we observe that the ratio of the amplitudes stabilizes at approximately  $\vev{|\varphi_b|}/\vev{|\varphi_a|}\sim 1/10$,
corresponding to the conversion of $O(1\%)$ of QCD axion number density $n_a$ into $n_b$. 
The increase in the frequency of $\varphi_b$ with time 
is due to the mass evolution $m_b \sim x^4$.
To  ease numerical integration (and to facilitate visualization of 
the oscillation pattern) in this example 
the oscillation frequency has been set to $\omega =50$.
We have verified that numerical integration  of 
\eqs{eq:oscillators2a}{eq:oscillators2b}  qualitatively   reproduces 
the features of the non-adiabatic LC in Fig.~\ref{fig:LC}.
%

%% file: Letter_PRLv3.bbl
%